\documentclass[lettersize,journal]{IEEEtran}
\usepackage{amsmath,amsfonts}
\usepackage{algorithmic}
\usepackage{algorithm}
\usepackage{array}
\usepackage[caption=false,font=normalsize,labelfont=sf,textfont=sf]{subfig}
\usepackage{textcomp}
\usepackage{stfloats}
\usepackage{url}
\usepackage{verbatim}
\usepackage{graphicx}
\usepackage{cite}
\usepackage{hyperref}
\usepackage{times}
\usepackage{latexsym}
\usepackage{amssymb}
\usepackage{pifont}
\usepackage{booktabs} 
\usepackage{threeparttable} 
\usepackage{graphicx} 
\usepackage{setspace}
\usepackage{multirow}
\usepackage[table,xcdraw]{xcolor}
\newcommand{\cmark}{\ding{51}} 
\newcommand{\xmark}{\ding{55}} 
\usepackage[T1]{fontenc}

\usepackage[utf8]{inputenc}

\usepackage{microtype}

\usepackage{inconsolata}

\usepackage{graphicx}

\usepackage{xcolor}

\definecolor{nightblue}{RGB}{25, 25, 112} 
\definecolor{orangered}{RGB}{255, 69, 0}  

\usepackage{amsmath,amsfonts}
\usepackage{algorithmic}
\usepackage{algorithm}

\begin{document}

\title{SecoustiCodec: Cross-Modal Aligned Streaming Single-Codecbook Speech Codec}

\author{Chunyu Qiang,
    Haoyu Wang,
    Cheng Gong,
    Tianrui Wang,
    Ruibo Fu, \emph{Member, IEEE,} 
    Tao Wang, 
    Ruilong Chen,
    Jiangyan Yi, \emph{Member, IEEE,}
    Zhengqi Wen, \emph{Member, IEEE,}
    Chen Zhang, 
    Longbiao Wang, \emph{Member, IEEE,} \\
    Jianwu Dang, \emph{Member, IEEE,}
    Jianhua Tao, \emph{Senior Member, IEEE}

\thanks{Chunyu Qiang is with School of New Media and Communication, Tianjin University, Tianjin, China, and also with Kuaishou Technology, Beijing, China. (e-mail: \url{qiangchunyu@tju.edu.cn})}
\thanks{Haoyu Wang, Cheng Gong, Tianrui Wang and Longbiao Wang are with Tianjin Key Laboratory of Cognitive Computing and Application, College of Intelligence and Computing, Tianjin University, Tianjin, China.  }
\thanks{Ruibo Fu and Tao Wang are with Institute of Automation Chinese Academy of Sciences, Beijing, China. }
\thanks{Ruilong Chen and Chen Zhang are with Kuaishou Technology, Beijing, China.}
\thanks{Jiangyan Yi, Zhengqi Wen, and Jianhua Tao are with Department of Automation, BNRist, Tsinghua University, Beijing, China.}
\thanks{Jianwu Dang is with Shenzhen Institute of Advanced Technology, Chinese Academy of Science, Guangdong, China.}
\thanks{Longbiao Wang is the corresponding author. (e-mail: \url{longbiao_wang@tju.edu.cn})}
}


\maketitle

\begin{abstract}
Speech codecs serve as a crucial bridge in unifying speech and text language models. Existing codec methods face several challenges in semantic encoding, such as residual paralinguistic information (e.g., timbre, emotion), insufficient semantic completeness, limited reconstruction capability, and lack of support for streaming.
To address these challenges, we propose SecoustiCodec, a cross-modal aligned low-bitrate streaming speech codec that disentangles semantic and paralinguistic information in a single-codebook space. 
To ensure semantic completeness and reconstruction fidelity, paralinguistic encoding is introduced to bridge the information gap between semantic and acoustic encoding. A semantic-only efficient quantization method based on VAE (Variational Autoencoder) and FSQ (Finite Scalar Quantization) is proposed. This approach alleviates the long-tail distribution problem of tokens while maintaining high codebook utilization. A semantic disentanglement method based on contrastive learning is proposed, which aligns text and speech in a joint multimodal frame-level space, effectively removing paralinguistic information from semantic encoding. An acoustic-constrained multi-stage optimization strategy is proposed to ensure robust and stable convergence. Figure~\ref{fig:pesq_kbps_below_2kbps} shows SecoustiCodec achieves SOTA (state-of-the-art) reconstruction quality (PESQ) of 1.77/2.58 at 0.27/1 kbps. The code and model weights for SecoustiCodec will be open-sourced upon the completion of the peer-review process. We've open-sourced SecoustiCodec's demo, code, and model weights at \url{https://qiangchunyu.github.io/SecoustiCodec_Page/}.
\end{abstract}

\begin{figure}[ht]
  \centering
  \includegraphics[width=\linewidth]{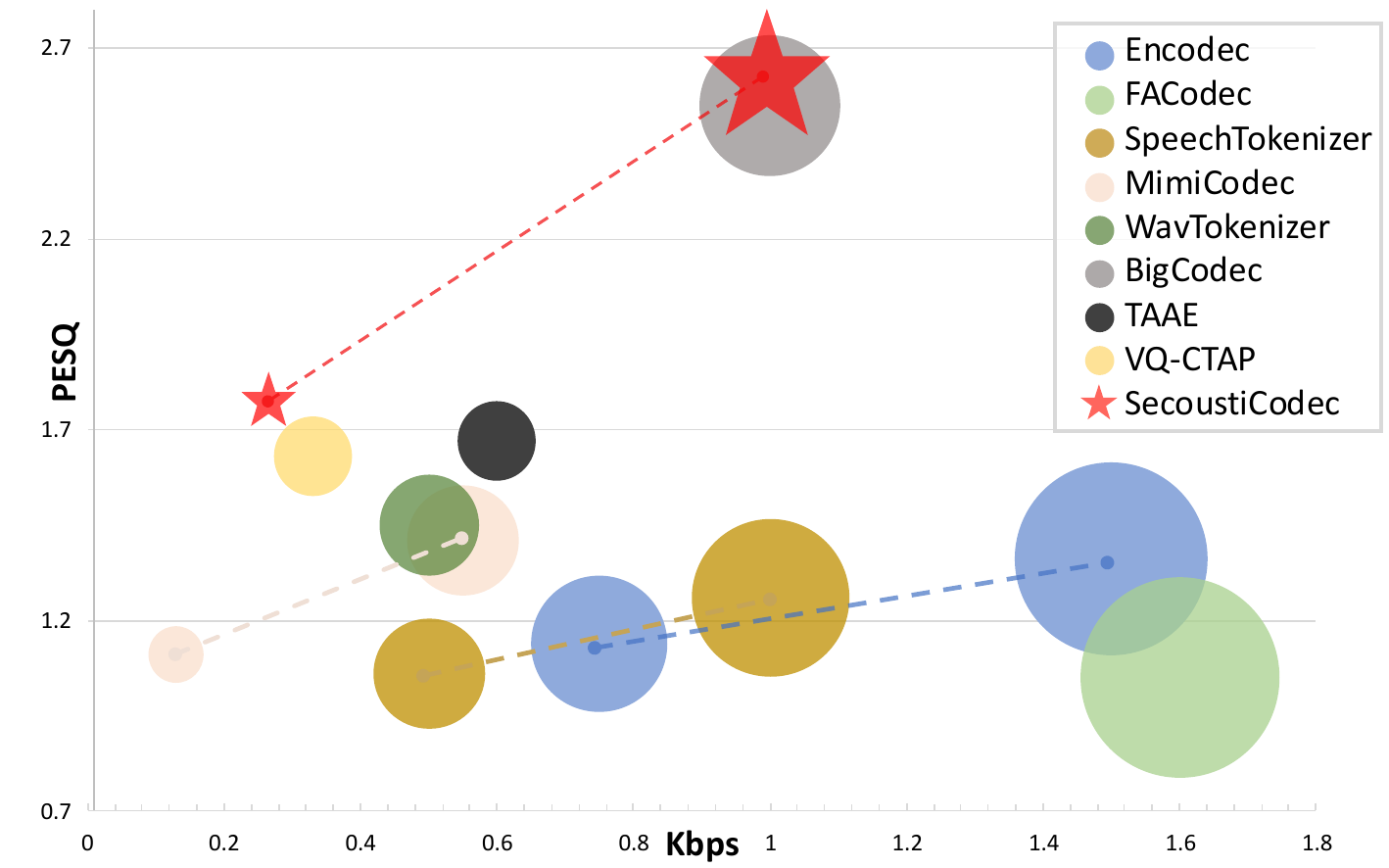}
  \caption{Comparison of different speech codecs operating below 2~kbps. The y-axis represents reconstruction quality (PESQ), while the x-axis indicates compression level (kbps). Circle sizes correspond to the number of discrete tokens encoded per second. SecoustiCodec supports streaming and claims SOTA performance in low-bitrate. Although BigCodec achieves comparable results, it neither supports causal streaming nor maintains parameter efficiency comparable to SecoustiCodec.}
  \label{fig:pesq_kbps_below_2kbps}
\end{figure}

\begin{IEEEkeywords}
Cross-Modal, Streaming Codec, Single-Codebook, Low-bitrate, Contrastive Learning
\end{IEEEkeywords}

\section{Introduction}
\label{sec:introduction}

\IEEEPARstart{L}{arge} language models (LLMs)\cite{zhao2023survey} have emerged as a focal point in generative AI research, driving advancements in text-to-speech (TTS)\cite{wang2023neural,zhang2023speak,kharitonov2023speak,qiang2024minimally,qiang2024high}, automatic speech recognition (ASR)\cite{bai2024seed}, and voice-based dialogue systems\cite{defossez2024moshi,zeng2024glm}. These applications critically depend on discrete speech representations, where neural codecs play a role analogous to text tokenizers in natural language processing (NLP). By converting complex speech waveforms into compact discrete tokens and modeling these tokens with LLMs, codecs enable unified training methodologies for both speech and text language models.

Current codecs aim to achieve two primary goals: high-fidelity speech reconstruction and semantic disentanglement. Residual Vector Quantization (RVQ)-based codecs\cite{zeghidour2021soundstream}, such as SoundStream\cite{zeghidour2021soundstream} and Encodec \cite{defossez2022high}, excel at acoustic fidelity through multi-codebook compression. However, directly training language models on the full set of acoustic tokens increases training complexity. To address this, models like VALL-E\cite{wang2023neural} predict the discrete tokens from the first quantizer using language models, followed by non-autoregressive models to predict the remaining tokens. While these methods treat the first quantizer as semantic encoding and the subsequent layers as acoustic encodings, the semantic encoding lacks genuine disentanglement capabilities. Removing paralinguistic information from semantic encoding significantly reduces the training complexity of language models\cite{borsos2022audiolm,spear-tts,defossez2024moshi}. The paralinguistic information refers to the gap between acoustic information and semantic information, encompassing elements such as timbre and emotion.

Models such as FACodec\cite{ju2024naturalspeech} and SpeechTokenizer\cite{zhang2023speechtokenizer} introduce supervised losses on the first quantizer to achieve semantic disentanglement. FACodec uses classification tasks and Gradient Reversal Layers (GRL) for attribute disentanglement, while SpeechTokenizer employs distillation from pre-trained models like HuBERT\cite{Hsu2021HuBERTSS} to disentangle semantic information. 
Multi-codebook codecs increase computational overhead and complicate modeling in downstream tasks. In response to these challenges, several single-codebook codec models have been developed, including WavTokenizer \cite{ji2024wavtokenizer}, BigCodec \cite{xin2024bigcodec}, and TAAE \cite{parker2024scaling}.
However, these methods fail to achieve high-quality reconstruction at low-bitrate and often adopt non-causal architectures, sacrificing streaming capability. 
Streaming capability is crucial for real-time interaction and low-latency scenarios in downstream tasks such as TTS, ASR, and voice-based dialogue systems.
MimiCodec\cite{defossez2024moshi}, similar to SpeechTokenizer, leverages pre-trained WavLM\cite{chen2022wavlm} for distillation-based semantic disentanglement while supporting streaming encoding and decoding through its causal architecture. Nevertheless, since representations from HuBERT/ WavLM inherently retain paralinguistic information, codecs relying on distillation from such pre-trained models fail to achieve true semantic disentanglement. In these models, the decoder reconstructs speech by directly summing semantic and residual encodings. This approach limits the disentanglement and reconstruction capabilities of semantic encodings. 

Contrastive learning has proven effective for cross-modal modeling. Compared to methods that use ASR loss\cite{du2024cosyvoice}, contrastive learning demonstrates a stronger ability to incorporate cross-modal textual information for semantic disentanglement\cite{qiang2024learning}.
In the field of text-audio cross-modal representation, CLIP-based\cite{radford2021learning} models have also been developed, including Wav2CLIP\cite{wu2022wav2clip}, AudioCLIP\cite{guzhov2022audioclip}, and CLAP\cite{elizalde2023clap}. However, their emphasis on global information limits their applicability to tasks requiring fine-grained frame-level tasks. 

\begin{figure}[ht]
  \centering
  \includegraphics[width=\linewidth]{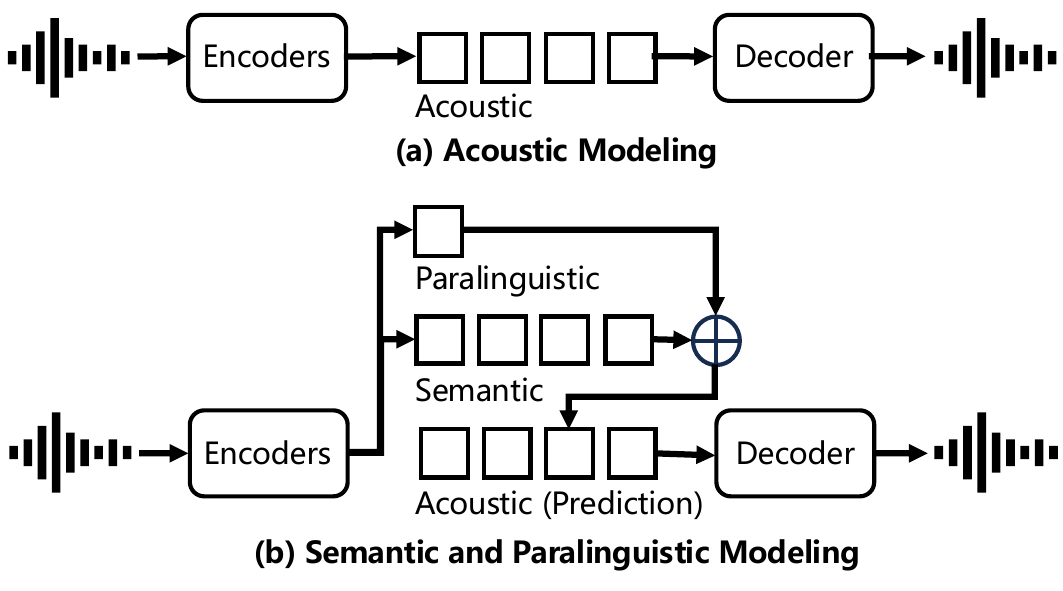}
  \caption{SecoustiCodec employs trained acoustic representations (frame-level continuous values) to constrain the joint training of semantic representations (frame-level discrete values) and paralinguistic representations (global-level continuous values). While we acknowledge that certain paralinguistic cues (e.g., nuanced emotional shifts) exhibit fine-grained variations not fully captured at a global level, we deliberately employ a global-level representation for paralinguistics. This design enables robust semantic decoupling while capturing dominant residual information between acoustic and semantic representations, such as speaker timbre and broad emotional characteristics. We posit that this global-level representation efficiently captures the majority of paralinguistic information while facilitating the core relationship:  $Semantic + Paralinguistic \approx Acoustic$}
  \label{fig:structure}
\end{figure}

To address these challenges, we propose SecoustiCodec, which independently models acoustic, semantic, and paralinguistic information to achieve low-bitrate high-fidelity real-time streaming encoding and decoding, as shown in Figure~\ref{fig:structure}. The contributions of this paper are as follows:  

1. We introduce a straightforward codec learning paradigm for effective semantic disentanglement. To ensure semantic completeness and reconstruction fidelity, paralinguistic encoding is introduced to bridge the information gap between semantic and acoustic encoding. 

2. We propos a semantic-only efficient quantization method based on VAE+FSQ. This approach alleviates the long-tail distribution problem of tokens while maintaining high codebook utilization. 

3. We propose a contrastive learning-based semantic disentanglement method, which aligns text and speech in a joint multimodal frame-level space. 

4. We design an acoustic-constrained multi-stage optimization strategy to ensure effective model convergence by gradually introducing modules and adjusting the influence of different loss components. Additionally, We adopt a causal architecture to support streaming encoding and decoding.

\section{Related work}
\label{sec:Related}

\subsection{Neural Codec}

Discretized vector quantization(VQ-VAE)\cite{van2017neural} has become a fundamental component for compression in codecs\cite{garbacea2019low}. The introduction of residual vector quantization (RVQ)\cite{zeghidour2021soundstream} enables high-quality audio compression with scalable bitrates. Current codecs are categorized into acoustic coding codecs (such as Encodec\cite{defossez2022high}, SoundStream\cite{zeghidour2021soundstream}, and SNAC\cite{siuzdak2024snac}) and semantic disentanglement codecs (such as FACodec\cite{ju2024naturalspeech}, SpeechTokenizer\cite{zhang2023speechtokenizer}, MimiCodec\cite{defossez2024moshi}, VQ-CTAP\cite{qiang2025vq} and SemantiCodec\cite{liu2024semanticodec}). Many single-codebook codec models are also proposed, such as WavTokenizer\cite{ji2024wavtokenizer}, BigCodec\cite{xin2024bigcodec} and TAAE\cite{parker2024scaling}.
Methods such as FACodec, SpeechTokenizer, and MimiCodec supervise the semantic encoding only in the first-dimensional VQ space, often incorporating ASR tasks\cite{ganin2015unsupervised} or distillation tasks from Hubert\cite{Hsu2021HuBERTSS}/ WavLM\cite{chen2022wavlm}. However, the representations from Hubert/ WavLM are not purely semantic and contain significant residual paralinguistic information. These models combine semantic encodings with other encodings during the decoding stage to reconstruct speech. This approach limits the disentanglement and reconstruction capability of the semantic encoding itself. 
In TTS tasks, there is no requirement for streaming in speech encoding. However, for voice-based dialogue tasks\cite{defossez2024moshi}, streaming speech encoding is essential. Among the methods mentioned, SoundStream, Encodec, and MimiCodec support streaming encoding and decoding through causal structures.

\subsection{Contrastive Learning}
Contrastive learning works by differentiating a target sample (positive) from distractor samples (negatives) based on an anchor representation. The objective is to maximize the similarity between the anchor and positive samples while minimizing the similarity between the anchor and negative samples. This approach has been widely applied in the field of computer vision, with notable examples such as Open AI's CLIP\cite{radford2021learning}, Florence\cite{yuan2021florence}, and ALIGN\cite{jia2021scaling}.
In the audio field, CLIP-based models have also been developed, including Wav2CLIP\cite{wu2022wav2clip}, AudioCLIP\cite{guzhov2022audioclip}, and CLAP\cite{elizalde2023clap}. These models focus on extracting global descriptive information from audio, with the primary goal of improving the performance of downstream audio classification tasks. The downstream tasks applied in the experimental parts of these studies are primarily speech classification tasks, such as sound event classification, instrument classification, acoustic scene classification, emotion recognition, keyword spotting, vocal sound classification, and speaker counting. Since the global features extracted by these methods lose temporal information, they cannot be converted back into frame-level acoustic features. VQ-CTAP\cite{qiang2025vq} is a frame-level speech representation model based on contrastive learning. However, it does not support streaming encoding and decoding.

\begin{figure*}[t]
 \centering
 \includegraphics[width=\linewidth]{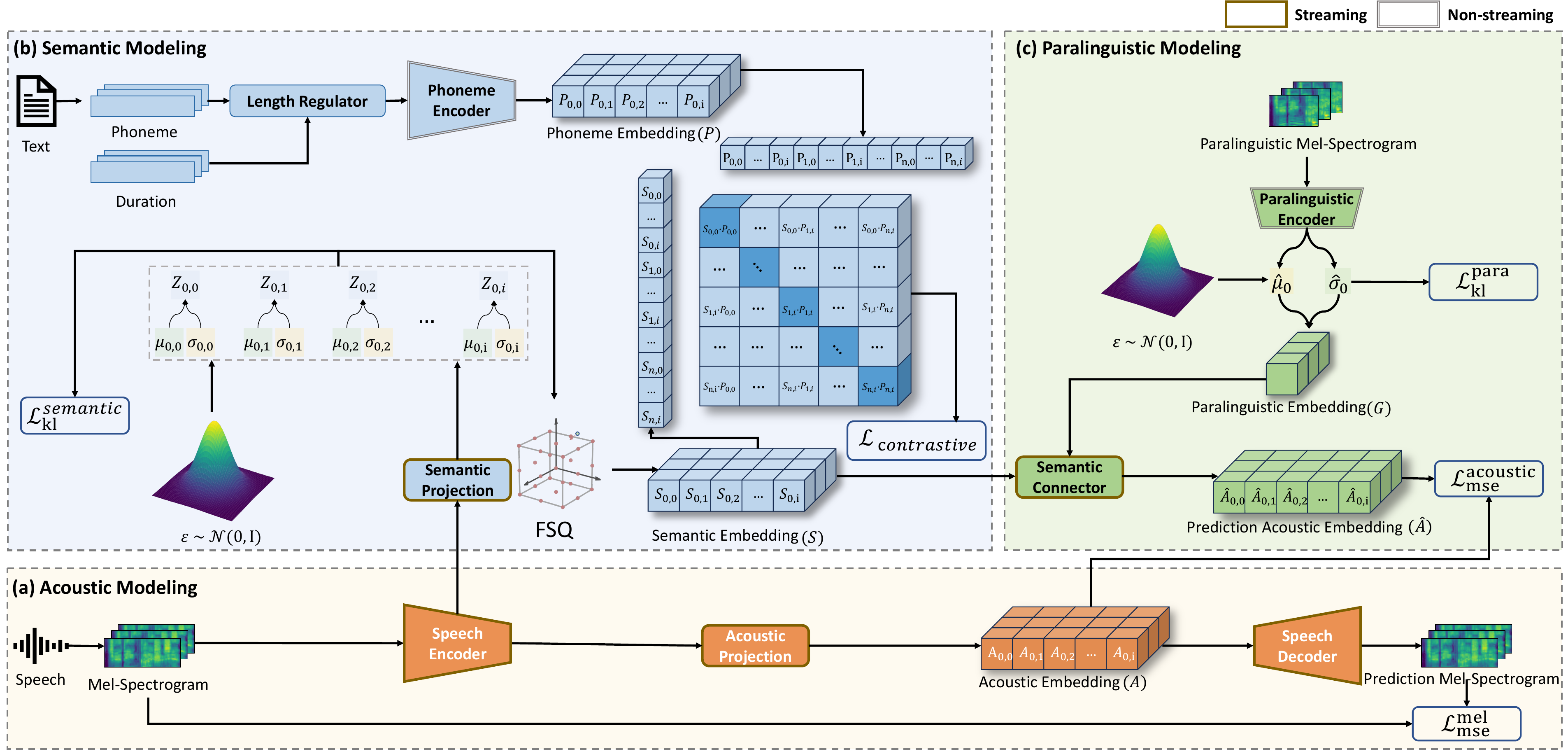}
 \caption{SecoustiCodec includes three modeling processes: (a) \textit{Acoustic Modeling}, (b) \textit{Semantic Modeling} and (c) \textit{Paralinguistic Modeling}. Modules outlined in \textcolor{orangered}{red} operate in a \textcolor{orangered}{streaming manner}, while those in \textcolor{nightblue}{blue} are \textcolor{nightblue}{non-streaming}. Phoneme embeddings $(P)$ are extracted from text, and target semantic embeddings $(S)$, acoustic embeddings $(A)$, and paralinguistic embeddings $(G)$ are extracted from speech. $(P)$ and $(S)$ are used to construct token-acoustic contrastive loss, which learns frame-level (dis)similarity between a batch of speech and text pairs. In the inference process, Acoustic Projection is not required; instead, semantic embedding and paralinguistic embedding are used to predict acoustic embedding. The mean values ($\mu$ and $\hat{\mu}$) from the VAE structure are directly used as inputs during inference, bypassing stochastic sampling. The name "SecoustiCodec" signifies a {\bf codec} that supports both {\bf se}mantic and a{\bf cousti}c encoding.}
 \label{fig:SecoustiCodec}
\end{figure*}

\section{Method}
\subsection{Overview}
As shown in Figure~\ref{fig:SecoustiCodec}, SecoustiCodec includes three modeling processes: (a) \textit{Acoustic Modeling}, which includes the speech encoder, acoustic projection, and speech decoder; (b) \textit{Semantic Modeling}, which comprises the semantic projection, phoneme encoder, VAE+FSQ, and contrastive learning module; and (c) \textit{Paralinguistic Modeling}, which contains the paralinguistic encoder, VAE, and semantic connector.
To enable streaming encoding and decoding during inference, the speech encoder, semantic projection, acoustic projection, semantic connector, and speech decoder are designed to be causal. A length regulator is employed to address the mismatch between the lengths of phoneme sequences and speech sequences. The phoneme and speech representations are concatenated in a joint multimodal space. A contrastive learning approach is used to learn this space by capturing the frame-level (dis)similarity between speech and phoneme pairs. We design a semantic-only efficient quantization method based on VAE+FSQ, which retains semantic content while further removing paralinguistic information from speech representations. 
SecoustiCodec extracts three independent representations: the single-codebook discrete semantic encoding $S$, the continuous acoustic encoding $A$, and the global-level paralinguistic encoding $G$. We anticipate that: $S + G \approx A$.

\subsection{Modeling of Acoustic, Semantic, and Paralinguistic}
$S_{in}$ denotes the input batch of speech data, $S_{in} \in \mathbb{R}^{B \times T_s \times D_s}$, where $B$ is the batch size, $T_s$ is the number of time frames, and $D_s$ is the number of spectral components (mel-spectrogram bands). 
$\tilde{P}_{in}$ denotes the input batch of phoneme data, $\tilde{P}_{in} \in \mathbb{R}^{B \times T_p \times D_p}$, where $T_p$ is the length of phoneme codes and $D_p$ is the dimensionality of the phoneme codes. During training, the ground-truth duration is used to extend the length of the phoneme sequences to $T_s$\cite{qiang2022back}. After the length regulator, the phoneme sequences become $P_{in} \in \mathbb{R}^{B \times T_s \times D_p}$, having the same length as the target speech sequences $S_{in}$. This duration-based alignment introduces limited misalignment between individual phonemes and speech frames. However, such mismatches occur only in a small fraction of training instances, resulting in negligible impact on the contrastive learning framework. 
The $P_{in}$ is then passed through a phoneme encoder. The $S_{in}$ is then passed through a speech encoder, a semantic projection, and an acoustic projection, respectively:
\begin{equation}
\begin{aligned}
P &= \mathrm{PhonemeEncoder}(P_{in})\\
S &= \mathrm{FSQ}(\mathrm{VAE}(\mathrm{Proj}(\mathrm{SpeechEncoder}(S_{in}))))\\
A &= \mathrm{Proj}(\mathrm{SpeechEncoder}(S_{in}))
\end{aligned}
\end{equation}

where ${P} \in \mathbb{R}^{B \times T_s/r \times d}$ are the phoneme representations, ${S} \in \mathbb{R}^{B \times T_s/r \times d}$ are the semantic representations, ${A} \in \mathbb{R}^{B \times T_s/4 \times d}$ are the acoustic representations. The phoneme encoder and speech encoder compress the lengths of the phoneme and speech representations by a factor of $r$, respectively. We bring ${S}$, and ${P}$ into a joint multimodal space of dimension $d$.
The paralinguistic speech $G_{in}$ serves as the input to the paralinguistic encoder:

\begin{equation}
\begin{aligned}
 {G} = \mathrm{ParalinguisticEncoder}(G_{in})
\end{aligned}
\end{equation}

$G$ denotes the paralinguistic embedding, where $G$ is a fixed-length global vector. This 3-second window is a hyperparameter setting based on empirical values from previous studies\cite{qiang2023improving}. $G \in \mathbb{R}^{B \times D_g}$, where $D_g$ is the number of dimensions. The paralinguistic encoder is a VAE-based model \cite{qiang2022style} that extracts paralinguistic information to address the one-to-many problem in decoding stage, such as timbre, style, and prosody, from the paralinguistic speech. In Equation (3), $D_{KL}$ refers to the KL loss, $\mathcal{N}(\cdot)$ represents a Gaussian distribution, and $({\hat{\mu}}, {\hat{\sigma}})$ denotes the $(mean, variance)$ of the paralinguistic latent space distribution. To address the KL collapse problem\cite{qiang2023improving}, a margin $\Delta$ is introduced to limit the minimum value of the KL loss as shown: 

\begin{equation}
\begin{aligned}
 \mathcal{L}_{kl}^{para} = max(0, D_{KL}[\mathcal{N}({\hat{\mu}},{\hat{\sigma}}^2)||\mathcal{N}(0, I)]-\Delta)
\end{aligned}
\end{equation}

To enable the semantic embedding $S$ to reconstruct the original speech, $S$ is fed into a semantic connector to predict the acoustic embedding $\hat{A}$. Since $S$ contains only semantic information, we provide $G$ as an input to supply paralinguistic information. The acoustic embedding $A$ is fed into the speech decoder to generate the prediction Mel-spectrogram $S_{p}$:

\begin{equation}
\begin{aligned}
 \hat{A} &= \mathrm{Semantic Connector}(S, G)\\
 S_{p} &= \mathrm{SpeechDecoder}(A)
\end{aligned}
\end{equation}

The mean squared error (MSE) loss measures the difference between the predicted acoustic embedding $\hat{A}$ and the ground-truth acoustic embedding $A$. Similarly, $S_{p}$ is compared with the ground-truth Mel-spectrogram $S_{in}$ to compute the reconstruction loss.

\begin{equation}
\begin{aligned}
\mathcal{L}_{mse}^{acoustic} &= MSE(\hat{A}, A)\\
\mathcal{L}_{mse}^{mel} &= MSE(S_{p}, S_{in})
\end{aligned}
\end{equation}

\subsection{Semantic-only Efficient Quantization}

As shown in Figure \ref{fig:SecoustiCodec} (b), a hybrid approach based on VAE+FSQ is proposed for vector quantization processing, which retains semantic content while further removing paralinguistic information from semantic encodings. In deep learning, a codebook refers to a collection of representative vectors or codes used to quantize data, typically in tasks like vector quantization or feature encoding. It helps reduce dimensionality, improve model efficiency, and enhance performance by mapping high-dimensional inputs to a finite set of meaningful representations\cite{van2017neural}. 

The input speech $S_{in}$ is processed by the speech encoder to generate an intermediate representation. This representation is then passed through the semantic projection to obtain a continuous and complete latent content space distribution. The semantic projection outputs the mean $\mu$ and variance $\sigma$ of a multivariate Gaussian distribution. Finally, a vector $z$ is sampled from this Gaussian distribution and used as the input for FSQ. Random operations in the network cannot be processed by backpropagation, "reparameterization trick" is introduced to VAE: $\boldsymbol{z} = {\mu} + {\sigma} \odot \phi ; \phi \sim \mathcal{N}(0, I)$

The calculation of KL Loss $\mathcal{L}_{kl}^{semantic}$for semantic projection is nearly identical to that of the paralinguistic encoder, as shown in Equation (3). However, it is important to note the difference in their outputs: the paralinguistic encoder produces $G$, a fixed-length global vector, while the semantic projection generates $z$, a variable-length frame-level vector.
The input $z$ undergoes quantization in the FSQ module:
\begin{equation}
\begin{aligned}
 S &= \mathrm{Proj_{up}}(\mathrm{round}(\lfloor L/2 \rfloor  \text{tanh}(\mathrm{Proj_{down}}(z))))
\end{aligned}
\end{equation}
where the function $f$ performs both dimensionality reduction and bounding. This process projects $z$ into a $d$-dimensional low-rank space, ensuring that each channel or element of $S$ takes one of $L$ unique values. The codebook is the implied codebook, given by the product of these per-channel codebook sets, with $codebooksize=L^d$. Finally, the quantized low-rank representation is projected back to its original dimensionality. The straight-through estimation is used to approximate the gradients of VAE and FSQ.

\subsection{Cross-Modal Token-Acoustic Contrastive Learning}

As shown in Figure \ref{fig:SecoustiCodec} (b), to extract frame-level representations, $S$ and $P$ within a batch are reshaped into 2D matrices $S_{re}$ and $P_{re}$, where $S_{re}$ and $P_{re} \in \mathbb{R}^{(B * T_s) \times d}$. This approach is beneficial for contrastive learning, as it increases the number of sample pairs per step, thereby enhancing the learning process.

With $S_{re}$ and $P_{re}$ now comparable, we can measure their similarity: $C = \tau*({S}_{re} \cdot {P}_{re}^\top)$

{\noindent}where $\tau$ is a temperature parameter used to scale the range of logits. The resulting similarity matrix $C \in \mathbb{R}^{(B * T_s) \times (B * T_s)}$ contains $(B * T_s)$ correct pairs along the diagonal and $(B * T_s)^2-(B * T_s)$ incorrect pairs in the off-diagonal elements. As the extracted intermediate representation includes contextual information, only the current frame corresponds to a positive sample.
The contrastive loss is calculated:

\begin{equation}
     \mathcal{L}_{contrastive}= 0.5 * (\ell_{speech}(C) + \ell_{phoneme}(C))
\end{equation}

{\noindent}where $\ell = \frac{1}{B * T_s}\sum_{i=0}^{B * T_s} \log diag (softmax(C))$


{\noindent}along the speech and phoneme axes, respectively. This symmetric cross-entropy loss ($\mathcal{L}_{contrastive}$) is computed over the similarity matrix to jointly train the speech encoder and the phoneme encoder, enabling the model to learn meaningful representations from both modalities simultaneously.

\subsection{Acoustic-Constrained Multi-Stage Optimization Strategy}
The acoustic-constrained multi-stage optimization strategy is designed to ensure effective model convergence by gradually introducing modules and adjusting the influence of various loss components, as detailed in Algorithm \ref{alg:training}. We employ the trained acoustic representation to impose constraints on the joint training process for semantic and paralinguistic representations. The training process involves the following loss functions: $\mathcal{L}_{kl}^{para}$, $\mathcal{L}_{kl}^{semantic}$, $\mathcal{L}_{mse}^{acoustic}$, $\mathcal{L}_{mse}^{mel}$, and $\mathcal{L}_{contrastive}$. The variable $step$ represents the current training step.

\textbf{Stage 1} (Acoustic Modeling) Algorithm \ref{alg:training}  (Lines 3-4): The focus is on learning acoustic representations.
During this phase, only the speech encoder, acoustic projection, and speech decoder modules are trained, with the loss function limited to $\mathcal{L}_{mse}^{mel}$. This targeted approach allows for the refinement of acoustic features before moving on to more complex tasks.

\textbf{Stage 2} (Semantic, and Paralinguistic Modeling) Algorithm \ref{alg:training} (Lines 5-13): The focus is on learning semantic representations.
Here, the weights of the speech encoder, acoustic projection, and speech decoder modules are frozen, while the phoneme encoder, paralinguistic encoder, semantic projection, and semantic connector are trained. Initially, the model uses $\mathcal{L}_{mse}^{acoustic}$ and $\mathcal{L}_{contrastive}$ for training. Notably, the weight $\beta$ of $\mathcal{L}_{contrastive}$ is significantly smaller than $\alpha$ of $\mathcal{L}_{mse}^{acoustic}$.
The KL terms are progressively introduced using linear warm-up schedules. When the $step$ exceeds the specified starting step $kl_{start}^{para}$ for $\mathcal{L}_{kl}^{para}$, this loss is added to the training process. Its weight gradually increases until $step$ surpasses the specified end step, at which point the weight is fixed at $kl_{upper}^{para}$. A similar process applies to $\mathcal{L}_{kl}^{semantic}$, with corresponding starting and fixed weights defined by $kl_{start}^{semantic}$ and $kl_{upper}^{semantic}$, respectively.

Algorithm \ref{alg:training} outlines the systematic introduction of different modules and loss functions, along with their corresponding weight adjustments. This optimization strategy is designed to enhance model convergence by strategically introducing and adjusting the impact of various loss components throughout the training process.

\begin{algorithm*}[t]
\caption{Acoustic-Constrained Multi-Stage Optimization}\label{alg:training}
\begin{algorithmic}
\setstretch{1.0} 
\small 
\STATE Initialize $step$, $stage1_{end}$, $kl_{start}^{para}$, $kl_{end}^{para}$, $kl_{upper}^{para}$, 
\STATE \hskip2em $kl_{start}^{semantic}$, $kl_{end}^{semantic}$, $kl_{upper}^{semantic}$, $\alpha$, $\beta$
\FOR{each training $step$}
    \IF{$step \leq stage1_{end}$}
        \STATE Train Speech Encoder, Acoustic Projection \& Speech Decoder with $\mathcal{L}_{mse}^{mel}$
    \ELSE
        \STATE Freeze modules from Stage 1
        \STATE $\mathcal{L}_{total} \gets \alpha\mathcal{L}_{mse}^{acoustic} + \beta\mathcal{L}_{contrastive}$
        \IF{$step > kl_{start}^{para}$}
            \STATE $\gamma \gets kl_{upper}^{para} \cdot \min(1, \frac{step - kl_{start}^{para}}{kl_{end}^{para} - kl_{start}^{para}})$
            \STATE $\mathcal{L}_{total} \gets \mathcal{L}_{total} + \gamma\mathcal{L}_{kl}^{para}$
        \ENDIF
        \IF{$step > kl_{start}^{semantic}$}
            \STATE $\delta \gets kl_{upper}^{semantic} \cdot \min(1, \frac{step - kl_{start}^{semantic}}{kl_{end}^{semantic} - kl_{start}^{semantic}})$
            \STATE $\mathcal{L}_{total} \gets \mathcal{L}_{total} + \delta\mathcal{L}_{kl}^{semantic}$
        \ENDIF
        \STATE Train Phoneme Encoder, Paralinguistic Encoder, Semantic Projector \& Connector with $\mathcal{L}_{total}$
    \ENDIF
\ENDFOR
\end{algorithmic}
\end{algorithm*}

\section{Experiments Procedures}

\subsection{Model Details}
\label{sec:ModelDetails}
Our architecture draws inspiration from MimiCodec\cite{defossez2024moshi}. All components employ causal operations to ensure streaming encoding and decoding.
The speech encoder/decoder is built using causal SeaNet encoder blocks \cite{tagliasacchi2020seanet}, the speech encoder consists of 80-channel convolutional layers with kernel size 7, ELU activation, and dilation base 2. The speech decoder mirrors this architecture with transposed convolutions for mel-spectrogram reconstruction. Both use 1 residual layer with true skip connections and a compression factor of 2.
The acoustic projection employs a causally masked transformer with 8 layers and 8-head attention, featuring RoPE positional encoding \cite{su2024roformer}. The 512-dimensional model includes layer normalization and a 2048-unit feedforward network, processing 250-frame context windows with convolutional layout integration.
The semantic projection combines a causal transformer (identical architecture to acoustic projection) with variational inference. Two linear projections ($\mu$ and $\sigma$) map the transformer's 512D output to latent variables using the reparameterization trick.
The semantic connector shares architectural parameters with the acoustic projection module, forming an 8-layer transformer with identical attention mechanisms and normalization schemes to maintain dimensional consistency across semantic representations.
The phoneme encoder employs a convolutional layer (ReLU-activated) followed by 4 transformer layers and linear projection. The paralinguistic encoder utilizes a VAE structure with 6 convolutional layers and SE-ResNet blocks \cite{hu2018squeeze}. All encoder outputs undergo layer normalization before fusion. The vocoder used in this experiment is HifiGAN \cite{kong2020hifi}.
For the multi-stage optimization strategy, the parameters are as follows: $stage1_{end}=1e4$, $kl_{start}^{semantic} = 2e4$, $kl_{start}^{para} = 2e4$, $kl_{end}^{semantic} = 3e4$, $kl_{end}^{para} = 3e4$, $kl_{upper}^{semantic} = 1e-5$, $kl_{upper}^{para} = 1e-5$, $\alpha = 1$, $\beta = 1e-5$. 
The model is trained using 8 NVIDIA TESLA A800 80GB GPUs, with a batch size of 64 per GPU. Adam\cite{Kingma2014AdamAM} is used as the optimizer, with an initial learning rate of 2e-4. The number of sample pairs per step ranges from 4,000 to 32,000 during the training process. 

\subsection{Datasets}

For the labeled text-speech paired data, we integrate our internal dataset with the AISHELL-3 dataset \cite{shi2020aishell} and the LibriTTS dataset \cite{zen2019libritts}, resulting in a combined total of 1,000 hours of recordings from 3,000 speakers.
All speech waveforms are sampled at 22kHz and converted to 80-band mel spectrograms with a window size of 1024 and a hop size of 256. 

\begin{table*}[t]
 \caption{Reconstruction Results of Different Codec Models}
 \label{tab:comparison}
 \centering
\resizebox{\linewidth}{!}{ 
\begin{tabular}{ll|ccccccccc
>{\columncolor[HTML]{ECF4FF}}c 
>{\columncolor[HTML]{ECF4FF}}c 
>{\columncolor[HTML]{ECF4FF}}c 
>{\columncolor[HTML]{ECF4FF}}c 
>{\columncolor[HTML]{ECF4FF}}c 
>{\columncolor[HTML]{ECF4FF}}c }
\hline
\multicolumn{2}{l|}{\textbf{Model}}                                              & GT               & \multicolumn{2}{c}{Encodec\cite{defossez2022high}} & \multicolumn{2}{c}{FACodec\cite{ju2024naturalspeech}}          & \multicolumn{2}{c}{SpeechTokenizer\cite{zhang2023speechtokenizer}}  & \multicolumn{2}{c}{MimiCodec\cite{defossez2024moshi}} & VQ-CTAP\cite{qiang2025vq}          & WavTokenizer\cite{ji2024wavtokenizer}     & BigCodec\cite{xin2024bigcodec}         & TAAE\cite{parker2024scaling}             & \multicolumn{2}{c}{\cellcolor[HTML]{ECF4FF}\textbf{SecoustiCodec}} \\ \hline
\multicolumn{2}{l|}{\textbf{Bitrate$\downarrow$}}                                & \textbackslash{} & 1.5kbps       & 6kbps       & 1.6kbps           & 4.8kbps          & 1kbps             & 4kbps            & 0.55kbps       & 4.4kbps      & 0.33kbps         & 0.5kbps          & 1kbps            & 0.6kbps          & 0.27kbps                      & 1kbps                              \\ \hline
\multicolumn{2}{l|}{\textbf{Frame Rate$\downarrow$}}                             & \textbackslash{} & \multicolumn{2}{c}{75Hz}    & \multicolumn{2}{c}{80Hz}             & \multicolumn{2}{c}{50Hz}             & \multicolumn{2}{c}{12.5Hz}    & 25Hz             & 40Hz             & 80Hz             & 25Hz             & 20Hz                          & 80Hz                               \\ \hline
\multicolumn{2}{l|}{\textbf{Nq$\downarrow$}}                                     & \textbackslash{} & 2             & 8           & 2                 & 6                & 2                 & 8                & 4              & 32           & 1                & 1                & 1                & 1                & \multicolumn{2}{c}{\cellcolor[HTML]{ECF4FF}1}                      \\ \hline
\multicolumn{2}{l|}{\textbf{Tokens/s$\downarrow$}}                               & \textbackslash{} & 150           & 600         & 160               & 480              & 100               & 400              & 100            & 800          & 25               & 40               & 80               & 25               & 20                            & 80                                 \\ \hline
\multicolumn{2}{l|}{\textbf{Param Size$\downarrow$}}                             & \textbackslash{} & \multicolumn{2}{c}{57MB}    & \multicolumn{2}{c}{406MB}            & \multicolumn{2}{c}{396MB}            & \multicolumn{2}{c}{303MB}     & 105MB            & 308MB            & 608MB            & 3,636MB          & \multicolumn{2}{c}{\cellcolor[HTML]{ECF4FF}379MB}                  \\ \hline
\multicolumn{2}{l|}{\textbf{Training   Dataset$\downarrow$}}                     & \textbackslash{} & \multicolumn{2}{c}{17,000h} & \multicolumn{2}{c}{500,000h}         & \multicolumn{2}{c}{1000h}            & \multicolumn{2}{c}{1000h}     & 11,000h          & 80,000h          & 1,000h           & 100,000h         & \multicolumn{2}{c}{\cellcolor[HTML]{ECF4FF}1,000h}                 \\ \hline
\multicolumn{2}{l|}{\textbf{Causal}}                                             & \textbackslash{} & \multicolumn{2}{c}{\cmark}  & \multicolumn{2}{c}{\xmark}           & \multicolumn{2}{c}{\xmark}           & \multicolumn{2}{c}{\cmark}    & \xmark           & \xmark           & \xmark           & \xmark           & \multicolumn{2}{c}{\cellcolor[HTML]{ECF4FF}\cmark}                 \\ \hline
\multicolumn{2}{l|}{\textbf{Latency$\downarrow$}}                                & \textbackslash{} & \multicolumn{2}{c}{13.39ms} & \multicolumn{2}{c}{\textbackslash{}} & \multicolumn{2}{c}{\textbackslash{}} & \multicolumn{2}{c}{84.4ms}    & \textbackslash{} & \textbackslash{} & \textbackslash{} & \textbackslash{} & \multicolumn{2}{c}{\cellcolor[HTML]{ECF4FF}12.08ms}                \\ \hline
\multicolumn{1}{l|}{}                                           & \textbf{Total} & \textbackslash{} & \multicolumn{2}{c}{0.004}   & \multicolumn{2}{c}{\textbackslash{}} & \multicolumn{2}{c}{\textbackslash{}} & \multicolumn{2}{c}{0.056}     & \textbackslash{} & \textbackslash{} & \textbackslash{} & \textbackslash{} & \multicolumn{2}{c}{\cellcolor[HTML]{ECF4FF}0.04}                   \\ \cline{2-17} 
\multicolumn{1}{l|}{}                                           & \textbf{Enc.}  & \textbackslash{} & \multicolumn{2}{c}{0.002}   & \multicolumn{2}{c}{\textbackslash{}} & \multicolumn{2}{c}{\textbackslash{}} & \multicolumn{2}{c}{0.033}     & \textbackslash{} & \textbackslash{} & \textbackslash{} & \textbackslash{} & \multicolumn{2}{c}{\cellcolor[HTML]{ECF4FF}0.002}                  \\ \cline{2-17} 
\multicolumn{1}{l|}{\multirow{-3}{*}{\textbf{RTF$\downarrow$}}} & \textbf{Dec.}  & \textbackslash{} & \multicolumn{2}{c}{0.002}   & \multicolumn{2}{c}{\textbackslash{}} & \multicolumn{2}{c}{\textbackslash{}} & \multicolumn{2}{c}{0.023}     & \textbackslash{} & \textbackslash{} & \textbackslash{} & \textbackslash{} & \multicolumn{2}{c}{\cellcolor[HTML]{ECF4FF}0.038}                  \\ \hline
\multicolumn{2}{l|}{\textbf{PESQ$\uparrow$}}                                     & 4.64             & 1.36          & 2.28        & 1.05              & 2.90             & 1.26              & 2.36             & 1.41           & 3.28         & 1.63             & 1.45             & 2.55             & 1.67             & 1.77                          & \textbf{2.58}                      \\ \hline
\multicolumn{2}{l|}{\textbf{Spk Sim$\uparrow$}}                                  & 1.00             & 0.9           & 0.98        & 0.80              & 0.98             & 0.74              & 0.97             & 0.85           & 0.97         & 0.89             & 0.86             & \textbf{0.97}    & 0.83             & 0.92                          & 0.95                               \\ \hline
\multicolumn{2}{l|}{\textbf{Emo Sim$\uparrow$}}                                  & 1.00             & 0.86          & 0.96        & 0.79              & 0.97             & 0.84              & 0.97             & 0.87           & 0.97         & 0.92             & 0.91             & 0.94    & 0.92             & 0.93                          & \textbf{0.97}                                \\ \hline
\multicolumn{2}{l|}{\textbf{LSD$\downarrow$}}                                    & 0.00             & 1.06          & 0.94        & 1.26              & 0.83             & 1.14              & 1.00             & 1.19           & 0.96         & 0.90             & 0.97             & 0.84             & 1.32             & 0.85                          & \textbf{0.77}                      \\ \hline
\multicolumn{2}{l|}{\textbf{MCD$\downarrow$}}                                    & 0.00             & 1.63          & 0.85        & 4.07              & 0.86             & 2.24              & 1.07             & 1.79           & 0.58         & 1.86             & 1.94             & \textbf{0.94}    & 2.59             & 1.72                          & 1.32                               \\ \hline
\multicolumn{2}{l|}{\textbf{MSEP$\downarrow$}}                                     & 0.00             & 36.03         & 17.06       & 630.71            & 12.15            & 40.83             & 18.93            & 43.28          & 9.10         & 29.00            & 39.29            & 18.26            & 28.57            & 39.66                         & \textbf{18.25}                     \\ \hline
\multicolumn{2}{l|}{\textbf{Mismatchrate$\downarrow$}}                           & 0.00             & 0.18          & 0.13        & 0.28              & 0.05             & 0.11              & 0.06             & 0.09           & 0.05         & 0.07             & 0.08             & 0.06             & 0.07             & 0.07                          & \textbf{0.05}                      \\ \hline
\multicolumn{2}{l|}{\textbf{WER$\downarrow$}}                                    & 3.05             & 5.60          & 3.37        & 15.61             & 3.47             & 16.97             & 3.62             & 10.69          & 3.23         & 17.06            & 19.70            & 4.10             & 30.24            & 11.58                         & \textbf{3.99}                      \\ \hline
\end{tabular}
}

\vspace{0.2in}

\begin{tablenotes}
\small
\item GT: {\bf G}round {\bf T}ruth Waveforms; Nq: {\bf N}umber of {\bf Q}uantizers; Causal: Streaming Support; RTF: {\bf R}eal {\bf T}ime {\bf F}actor; \cmark: Enabled; \xmark: Disabled.
\item  Best results from models with a single quantizer (hence directly comparable to SecoustiCodec) are in {\bf bold}.
\end{tablenotes}

\end{table*}

\subsection{Compared Method and Tasks}
\label{sec:ComparedMethod}
To evaluate the performance of the proposed model, we conduct experiments on speech reconstruction task. We compare {\bf SecoustiCodec} with eight representative Codec models, as summarized in Table \ref{tab:comparison}. The table lists key attributes of each model, including bitrate, frame rate, number of quantizers, tokens per
second, parameter size, training data duration, and whether the model employs a causal structure supporting streaming encoding and decoding. The comparison includes classic streaming Codec models like {\bf Encodec}\footnote{https://github.com/facebookresearch/encodec}, semantic disentanglement models like {\bf SpeechTokenizer}\footnote{https://github.com/ZhangXInFD/SpeechTokenizer} and {\bf MimiCodec} (Moshi)\footnote{https://huggingface.co/kyutai/mimi}, as well as {\bf FACodec} (NaturalSpeech 3)\footnote{https://github.com/lifeiteng/naturalspeech3\_facodec}, which disentangles attributes through supervised tasks, and single-codebook models like {\bf VQ-CTAP}, {\bf WavTokenizer}\footnote{https://github.com/jishengpeng/WavTokenizer}, {\bf BigCodec}\footnote{https://github.com/Aria-K-Alethia/BigCodec}, and {\bf TAAE}\footnote{https://github.com/Stability-AI/stable-codec}.

We also compare multi-codebook models under different bitrate conditions to demonstrate the effectiveness of SecoustiCodec. Despite SecoustiCodec’s denser structure, it does not increase the parameter count. Additionally, as shown in Table \ref{tab:Ablation}, we perform extensive ablation studies to analyze various factors. These include the quantization method, the use of a causal structure for streaming support, the inclusion of pitch features alongside Mel spectrograms, the application of a multi-stage optimization strategy, and the dimensionality of the acoustic embedding. Finally, we also evaluate the performance of SecoustiCodec in single-speaker scenarios and Voice Conversion (VC) task.

\subsection{Evaluation Metrics}
We evaluate the model using several objective metrics to ensure comprehensive performance assessment. These metrics include: PESQ (Perceptual Evaluation of Speech Quality)\footnote{\url{https://github.com/ludlows/PESQ}},
Speaker Similarity\footnote{\url{https://github.com/resemble-ai/Resemblyzer}},
Emotion Similarity\footnote{\url{https://huggingface.co/emotion2vec}},
LSD (Log-Spectral Distance),
MCD (Mel-Cepstral Distortion),
MSEP (Mean Squared Error of Pitch),
MR (Voiced/unvoiced Mismatch Rate),
NISQA (Speech Quality and Naturalness Assessment)\footnote{\url{https://github.com/gabrielmittag/NISQA}}, and
WER (Word Error Rate)\footnote{\url{https://github.com/modelscope/FunASR}}.

\section{Results and Analysis}

\begin{table*}[]
 \caption{Ablation Studies of SecoustiCodec}
 \label{tab:Ablation}
 \centering
\resizebox{\linewidth}{!}{ 
\begin{tabular}{l|ccccc|ccccccccc}
\hline
\textbf{Model}                   & \textbf{Quant} & \textbf{Causal} & \textbf{F0} & \textbf{Stage} & \textbf{Acous-Dim} & \textbf{PESQ$\uparrow$} & \textbf{Spk   Sim$\uparrow$} & \textbf{Emo   Sim$\uparrow$} & \textbf{LSD$\downarrow$} & \textbf{MCD$\downarrow$} & \textbf{MSEP$\downarrow$} & \textbf{MR$\downarrow$} & \textbf{NISQA$\uparrow$} & \textbf{WER$\downarrow$} \\ \hline
Secousti-VQ                      & VQ-VAE         & \cmark          & \xmark      & \cmark         & 256                & 1.28                    & 0.87                         & 0.92                         & 0.96                     & 2.31                     & 54.54                     & 0.09                    & 2.72                   & 47.28                    \\ \hline
Secousti-VQ w/o Stage            & VQ-VAE         & \cmark          & \xmark      & \xmark         & 256                & 1.20                    & 0.87                         & 0.91                         & 0.98                     & 2.56                     & 71.89                     & 0.11                    & 2.66                   & 56.28                    \\ \hline
Secousti-VQ w/o Causal           & VQ-VAE         & \xmark          & \xmark      & \cmark         & 256                & 1.44                    & 0.62                         & 0.8                          & 1.69                     & 5.42                     & 46.47                     & 0.18                    & 2.51                   & 14.31                    \\ \hline
Secousti-VQ w/o Causal w/o Stage & VQ-VAE         & \xmark          & \xmark      & \xmark         & 256                & 1.10                    & 0.74                         & 0.84                         & 1.03                     & 3.19                     & 155.64                    & 0.15                    & 1.86                   & 66.77                    \\ \hline
Secousti-VQ w/ F0                & VQ-VAE         & \cmark          & \cmark      & \cmark         & 256                & 1.02                    & 0.54                         & 0.71                         & 1.30                     & 4.27                     & 1286.17                   & 0.52                    & 0.97                   & 98.38                    \\ \hline
Secousti-VQ w/ F0 w/o Stage      & VQ-VAE         & \cmark          & \cmark      & \xmark         & 256                & 1.03                    & 0.58                         & 0.67                         & 1.50                     & 4.28                     & 519.01                    & 0.44                    & 1.35                   & 100.00                   \\ \hline
Secousti-SimVQ                   & SimVQ          & \cmark          & \xmark      & \cmark         & 256                & 1.65                    & 0.90                         & 0.92                         & 0.89                     & 1.84                     & \textbf{28.63}            & \textbf{0.07}           & 3.07                   & 16.95                    \\ \hline
Secousti-FSQ-8dim                & FSQ            & \cmark          & \xmark      & \cmark         & 8                  & 1.53                    & 0.93                         & 0.94                         & 0.89                     & 1.99                     & 48.53                     & \textbf{0.07}           & \textbf{4.07}          & 14.00                    \\ \hline
Secousti-FSQ-64dim               & FSQ            & \cmark          & \xmark      & \cmark         & 64                 & 1.57                    & 0.90                         & 0.91                         & 0.88                     & 1.89                     & 54.18                     & 0.09                    & 3.23                   & 18.67                    \\ \hline
Secousti-FSQ-256dim              & FSQ            & \cmark          & \xmark      & \cmark         & 256                & 1.71                    & 0.90                         & 0.92                         & 0.88                     & 1.78                     & 30.16                     & \textbf{0.07}           & 3.13                   & 15.13                    \\ \hline
Secousti-FSQ-64dim w/o Stage     & FSQ            & \cmark          & \xmark      & \xmark         & 64                 & 1.65                    & 0.91                         & 0.92                         & 0.87                     & 1.82                     & 40.18                     & 0.08                    & 3.18                   & 13.23                    \\ \hline
Secousti-VAE-FSQ-64dim           & VAE+FSQ        & \cmark          & \xmark      & \cmark         & 64                 & 1.60                    & 0.90                         & 0.92                         & 0.87                     & 1.86                     & 49.78                     & 0.08                    & 3.23                   & 17.04                    \\ \hline
\textbf{SecoustiCodec(0.27kbps)} & VAE+FSQ        & \cmark          & \xmark      & \cmark         & 256                & \textbf{1.77}           & \textbf{0.92}                & \textbf{0.93}                & \textbf{0.85}            & \textbf{1.72}            & 39.66                     & \textbf{0.07}           & 3.50                   & \textbf{11.58}           \\ \hline
\end{tabular}
}

\begin{tablenotes}
\small
\item {\bf Quant}: Quantization Method; {\bf Causal}: Streaming Support; {\bf F0}: Pitch Feature; {\bf Stage}: Multi-stage Training; {\bf Acous-Dim}: Acoustic Embedding Dimension.
\item \cmark: Enabled, \xmark: Disabled. Bold row indicates our standard configuration.
\end{tablenotes}

\end{table*}

\begin{table*}[b]
 \caption{Results of Single Speaker}
 \label{tab:Single}
 \centering
 \vspace{-0.1in}
\resizebox{0.7\linewidth}{!}{ 
\begin{tabular}{l|l|cccccccc}
\hline
\textbf{Model}                                                                             & \textbf{Bitrate} & \textbf{PESQ$\uparrow$} & \textbf{Spk Sim$\uparrow$} & \textbf{Emo Sim$\uparrow$} & \textbf{LSD$\downarrow$} & \textbf{MCD$\downarrow$} & \multicolumn{1}{l}{\textbf{F0$\downarrow$}} & \multicolumn{1}{l}{\textbf{MR$\downarrow$}} & \multicolumn{1}{l}{\textbf{WER$\downarrow$}} \\ \hline
\multirow{2}{*}{\begin{tabular}[c]{@{}l@{}}SecoustiCodec\\ ft Single Speaker\end{tabular}} & 0.27             & 2.50                    & 0.96                       & 0.80                       & 0.76                     & 1.18                     & 20.88                                       & 0.06                                        & 6.51                                         \\ \cline{2-10} 
                                                                                           & 1                & 3.51                    & 0.97                       & 0.97                       & 0.71                     & 0.89                     & 8.99                                        & 0.04                                        & 3.37                                         \\ \hline
\end{tabular}
}
\vspace{-0.2in}
\end{table*}

\begin{figure}[h]
 \centering
 \includegraphics[width=\linewidth]{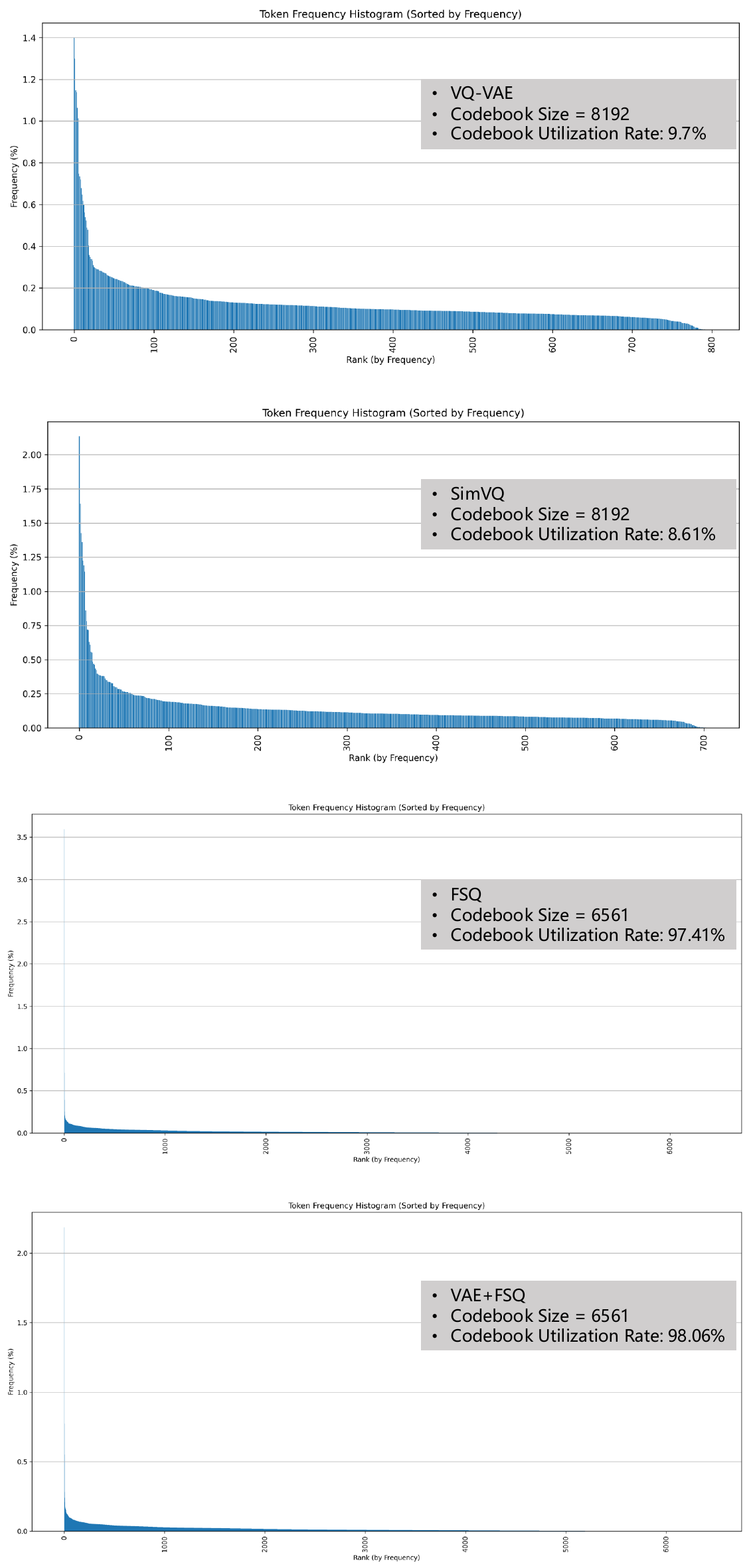}
 \caption{Codebook Utilization.}
 \label{fig:token_distribution}
\end{figure}

\begin{figure*}[h]
 \centering
 \includegraphics[width=0.85\linewidth]{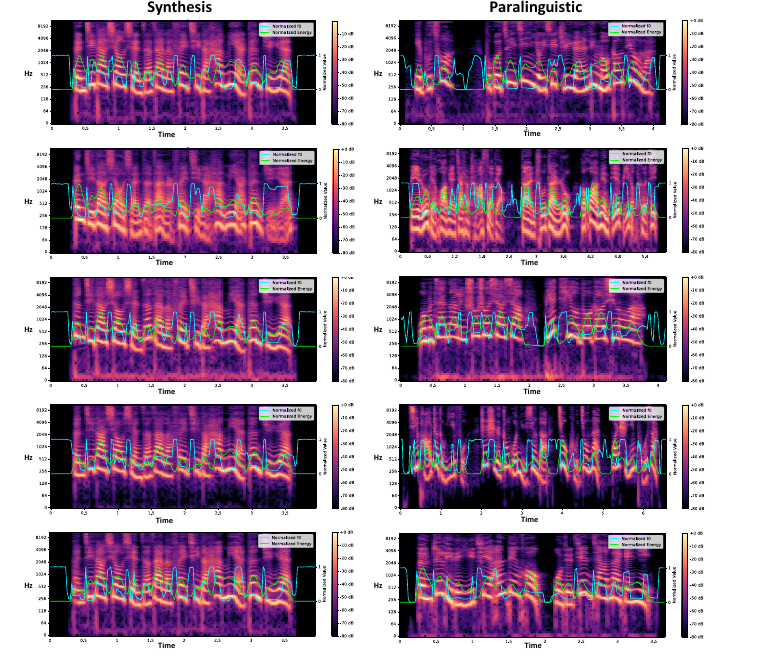}
 \caption{The Spectrograms, F0, and Energy of synthesized speech (the same semantic coding combining different paralinguistics). The bottom row is the ground-truth. The synthesized speech and the paralinguistic speech exhibit consistency in the numerical range and variation trends of spectrogram, F0, and energy.}
 \vspace{-0.3in}
 \label{fig:mel_f0_energy}
\end{figure*}

\subsection{Structure Comparison}

Table~\ref{tab:comparison} compares the structure of SecoustiCodec with mainstream codecs. SecoustiCodec achieves a low bitrate of 0.27kbps, a frame rate of 20Hz and a token rate of 20 tokens/s, second only to MimiCodec. In terms of quantization methods, it adopts the proposed VAE-FSQ approach instead of the RVQ or VQ structures used by other models. This enables single-codebook semantic encoding while supporting continuous-value acoustic encoding. Although the model has a relatively high parameter count, it remains smaller than FACodec, SpeechTokenizer, BigCodec and TAAE. Regarding training data, SecoustiCodec uses 1,000 hours of data, similar to SpeechTokenizer, MimiCodec and BigCodec, but significantly less than FACodec’s 50,000 hours and TAAE's 100,000 hours. Additionally, it employs a causal structure to support streaming encoding and decoding. Experimental results demonstrate that SecoustiCodec achieves optimal performance under conditions of causal structure, single-codebook encoding, low bitrate, and limited training data. In Table \ref{tab:comparison} , we present the initial latency and real-time factors (RTF) for all causal structure-supported streaming models. The real-time factor is defined as the ratio between the processing time and the speech duration, so that it is less than one when the method is faster than real time. We profiled all models on a single thread of a V800 GPU. SecoustiCodec exhibits an initial latency of 12.08 ms, which is lower than that of Encodec and MimiCodec. In terms of real-time factor, Encodec performs significantly better than SecoustiCodec. Notably, both Encodec and SecoustiCodec have an encoding RTF of 0.002. However, the decoding process for SecoustiCodec includes time-consuming vocoder operations, which increases its decoding real-time factor. Excluding the vocoder processing time, SecoustiCodec achieves a decoding real-time factor of 0.001, which is lower than that of Encodec.

\subsection{Evaluation of Speech Reconstruction}
Table~\ref{tab:comparison} presents the results of the speech reconstruction task. To demonstrate the effectiveness of SecoustiCodec, we compare its performance with other single-codebook methods and multi-codebook methods under different bitrate. 

SecoustiCodec(1kbps) achieves the best performance across multiple metrics, including PESQ, EmoSim, LSD, F0, MR and WER, surpassing some models even in scenarios where they use multiple codebooks. SecoustiCodec(0.27kbps 20Hz) is the model with the lowest bitrate and frame rate among single-codebook models. It achieves the best results across multiple metrics, including PSEQ, SpkSim, EmoSim, LSD, and WER, when compared to other single-codebook models, with the exception of BigCodec(1kbps 80Hz). SecoustiCodec(0.27kbps 20tokens/s) also outperforms multi-codebook models such as Encodec (1.5kbps 150tokens/s), FACodec (1.6kbps 160tokens/s), SpeechTokenizer (1kbps 100tokens/s), and MimiCodec (0.55kbps 100tokens/s), despite these models having a significantly higher token rate than SecoustiCodec.

Methods such as FACodec, SpeechTokenizer, and MimiCodec supervise the semantic encoding only in the first-dimensional VQ space, often incorporating ASR tasks or distillation tasks from Hubert\cite{Hsu2021HuBERTSS}/ WavLM\cite{chen2022wavlm}. However, the representations from Hubert/ WavLM are not purely semantic and contain significant residual paralinguistic information. These models combine semantic encodings with other encodings during the decoding stage to reconstruct speech. This approach limits the disentanglement and reconstruction capability of the semantic encoding itself. During training, the model relies on the full set of encodings to extract acoustic information, which prevents the semantic encoding from independently ensuring semantic completeness. As a result, substantial paralinguistic information often remains in the semantic encoding.
In contrast, SecoustiCodec benefits from its disentanglement of semantic encoding and its explicit modeling of paralinguistic information. This enables it to outperform other methods significantly in both  SpkSim and EmoSim. During training, SecoustiCodec independently models semantic, acoustic, and paralinguistic information. Additionally, it incorporates a task where the semantic encoding predicts the acoustic encoding, using paralinguistic encoding as input to bridge the information gap between the two. This approach helps the model learn pure semantic and paralinguistic representations while ensuring semantic completeness and robust reconstruction capabilities. Consequently, SecoustiCodec achieves superior results in reconstruction metrics such as PESQ, EmoSim, LSD, F0, MR, and WER. It is important to note that these results are obtained under challenging conditions, including the use of a causal structure, single-codebook encoding, high compression rates, and limited training data.

\subsection{Ablation Studies}
\label{sec:Ablation}

As shown in Table \ref{tab:Ablation}, we conduct extensive ablation experiments to validate various aspects of our model.
Regarding quantization methods, SecoustiCodec outperforms other models, such as Secousti-SimVQ, Secousti-VQ, and Secousti-FSQ, across all metrics except MSEP, demonstrating the superiority of the VAE-FSQ quantization method. We also compare the codebook utilization of different quantization methods, as illustrated in Figure \ref{fig:token_distribution}. VAE-FSQ achieves the highest codebook utilization (98.06\%), with nearly all tokens activated, significantly surpassing VQ-VAE and SimVQ. Additionally, VAE-FSQ minimizes the long-tail effect, with most token frequencies below 0.2\%, which aids language model training in TTS and voice dialogue tasks.
In terms of causal structure, Secousti-VQ without causal structure shows improvements in metrics like PESQ, MSEP, and WER compared to Secousti-VQ, though the causal structure variant performs better in other metrics. 
Regarding pitch features, adding pitch features to Secousti-VQ (Secousti-VQ w/ F0) negatively impacts nearly all metrics, particularly MSEP and WER, where it records the worst values among all ablation models. While pitch features aid acoustic modeling, they significantly interfere with semantic modeling capabilities.
Concerning the multi-stage training strategy, models that do not employ this strategy, such as Secousti-VQ w/o Stage, Secousti-VQ w/o Causal w/o Stage, Secousti-VQ w/ F0 w/o Stage, and Secousti-FSQ-64dim w/o Stage, show significant degradation in all metrics. This is because introducing additional loss terms during the acoustic modeling stage affects the convergence of the $\mathcal{L}_{mse}^{mel}$, ultimately reducing the representational capacity of the generated acoustic embeddings. The presence of the $\mathcal{L}_{mse}^{acoustic}$ further diminishes the semantic encoding reconstruction ability.
Regarding the dimensionality of acoustic embedding, we perform ablation experiments on the Secousti-FSQ structure. Since semantic encoding is mapped to an 8-dimensional embedding via a linear layer before FSQ, we aim for consistent dimensionality in acoustic encoding to reduce modeling complexity. However, we find that the dimensionality of the acoustic embedding significantly influences the model's overall capability. As evidenced by Secousti-FSQ-8dim, Secousti-FSQ-64dim, and Secousti-FSQ-256dim, increasing the acoustic embedding dimension improves all metrics. Ultimately, we select 256 as the acoustic encoding dimension for SecoustiCodec.

\begin{table}[h]
 \caption{Results of VC}
 \label{tab:VC}
 \centering
 \vspace{-0.1in}
\resizebox{\linewidth}{!}{ 
\begin{tabular}{llll}
\hline
\multicolumn{1}{c}{\textbf{Model}} & \multicolumn{1}{c}{\textbf{SpkSim↑}} & \multicolumn{1}{c}{\textbf{EmoSim↑}} & \multicolumn{1}{c}{\textbf{WER↓}} \\ \hline
SecoustiCodec-ASR-Loss             & 0.64                                 & 0.69                                 & \textbf{10.51}                    \\ \hline
\textbf{SecoustiCodec}             & \textbf{0.71}                        & \textbf{0.86}                        & 11.58                             \\ \hline
\end{tabular}
}
\vspace{-0.2in}
\end{table}

\subsection{Semantic-Paralinguistic Disentangle and Single-Speaker}
\label{sec:Disentangle}

We have added frame-level ablation experiments for ASR loss. For fairness, we replaced frame-level contrastive learning with frame-level phoneme classification loss (Secousti-ASR-Loss). To demonstrate the effect of semantic decoupling, we validated this on the VC task (replacing Paralinguistic Embedding at the decoder stage). As shown in Table \ref{tab:VC}, compared to ASR loss, frame-level contrastive loss achieved higher speaker similarity and emotional consistency, but Secousti-ASR performed better on the WER metric. Compared to methods that use ASR loss, frame-level contrastive learning demonstrates a stronger ability to incorporate cross-modal textual information for semantic disentanglement. Figure~\ref{fig:mel_f0_energy} illustrates the spectrograms, F0, and energy of speech generated by SecoustiCodec using the same semantic encoding combined with paralinguistic embeddings from different speakers. As shown in the figure, the synthesized speech exhibits strong alignment with the paralinguistic speech across multiple frequency domains, F0, and energy spectrograms. This is attributed to the disentanglement of paralinguistic information in the semantic encoding and the paralinguistic modeling capability of the paralinguistic encoder in SecoustiCodec. 

We conduct additional experiments in a single-speaker scenario. In tasks such as TTS and voice-based dialogue, it is often sufficient to use the timbre of a single target speaker. In practical applications, a decoder specifically trained for the target speaker is commonly used to enhance audio quality, speaker consistency, and expressiveness. As shown in Table \ref{tab:Single}, the fine-tuned single-speaker model achieves high performance across all metrics.

\section{Conclusions and Future Work}

This paper presents SecoustiCodec, a cross-modal aligned low-bitrate streaming speech codec designed for semantic disentanglement and high-fidelity real-time encoding and decoding in a single-codebook space.
The main contributions of this work include the independent modeling of acoustic, semantic, and paralinguistic information. During the training phase, paralinguistic encodings bridge the gap between semantic and acoustic encodings, ensuring both semantic completeness and reconstruction fidelity. The semantic-only efficient quantization method, based on VAE+FSQ, enhances codebook utilization (98.06\%) and addresses the long-tail distribution problem. Our semantic disentanglement approach, utilizing contrastive learning, aligns text and speech in a joint multimodal frame-level space, effectively removing paralinguistic information from semantic encoding. Furthermore, an acoustic-constrained multi-stage optimization strategy is employed to ensure effective model convergence, gradually introducing modules and adjusting the influence of different loss components. Experimental results show that SecoustiCodec achieves SOTA reconstruction quality at bitrates of 0.27 kbps and 1 kbps

In future work, we plan to explore unsupervised disentanglement methods to reduce dependence on labeled text data. Furthermore, experiments are conducted only on English and Chinese datasets, leaving the adaptability of the model to other languages unverified. We aim to address these limitations in future research to enhance the effectiveness and versatility of the model.

\bibliographystyle{IEEEtran}
\bibliography{trans}

\vfill

\end{document}